\documentclass[11pt]{article}
\begin{document}

\title{\bf Behavior of Time-varying Constants in Relativity}

\author{M. Sharif \thanks{msharif@math.pu.edu.pk} and H. Rizwana
Kausar\thanks{rizwa\_math@yahoo.com}\\
Department of Mathematics, University of the Punjab,\\
Quaid-e-Azam Campus, Lahore-54590, Pakistan.}

\date{}

\maketitle

In this paper, we consider Bianchi type III and Kantowski-Sachs
spacetimes and discuss the behavior of time-varying constants $G$
and $\Lambda$ by using two symmetric techniques, namely, kinematic
self-similarity and matter collineation. In the kinematic
self-similarity technique, we investigate the behavior of the first
and the second kinds. In the matter collineation technique, we consider
usual, modified, and completely modified matter collineation
equations while studying the behavior of these constants. Further,
we reduce the results for dust, radiation, and stiff fluids. We find
that $\Lambda$ is a decreasing time function while $G$ is an
increasing time function. This corresponds to the earlier results
available in the literature for other spacetimes. Further, we find
that the deceleration parameter attains a negative value, which shows
that the expansion of the universe is accelerating.


\newpage
\begin{center}
\section*{I. INTRODUCTION}
\end{center}

The time-varying behavior of the gravitational and the cosmological
constants, i.e., $G$ and $\Lambda$, have been among the most
controversial issues in cosmology. Some authors reveal the idea that
both $G$ and $\Lambda$ can be considered as non-constant, i.e., $G=
G(t)$ and $\Lambda= \Lambda(t)$, coupling scalars, while solving the
Einstein field equations (EFEs). The use of $G$ and $\Lambda$ in the
EFEs plays a significant role in cosmological, astronomical, and
quantum phenomena.

Improved astronomical techniques indicate that a non-zero value of
$\Lambda$ is required to discuss distant supernova. Analysis of high
red-shift supernova shows that the universe may be accelerating
[1-3] due to the presence of some type of vacuum energy. Some people
relate this vacuum energy density to a non-zero cosmological
constant. This indicates that the cosmological constant is a cause
of the expansion of the universe and plays an important role in the
evolution of the universe. In gravitational collapse, the
cosmological constant slows it down, which limits the size of a
black hole [4]. However, there is a fundamental problem related with
the value of $\Lambda$, which is assumed to be very small. This
value is about $120$ orders less than the magnitude of the vacuum
energy-density calculated in quantum field theory. A phenomological
solution to this problem is suggested by considering $\Lambda$ as a
function of time. In the early universe, $\Lambda$ was large, but
decreased with the expansion of the universe and the creation of
photons [5].

The gravitational attraction, explained as a result of the curvature
of spacetime, is proportional to $G$. According to quantum theory,
the non-homogeneity of the gravitational field causes $G$ to change
rapidly in small intervals of time [6]. The EFEs with $G$ and
$\Lambda$ are given by
\begin{equation}\label{1}
R_{ab}-\frac{1}{2}R g_{ab}-\Lambda(t) g_{ab}=\frac{8\pi G(t)}{c^4}
T_{ab}, \quad\ (a,b=0,1,2,3),
\end{equation}
where $R_{ab}$, $g_{ab}$, and $T_{ab}$ represent the components of
the Ricci, metric, and energy-momentum tensors respectively, and $R$
is the Ricci scalar.

The time variation of $G$ was originally raised by Dirac [7]. He
proposed that the gravitational constant varied with age of the
universe. Modern theories, like string theory and Brans-Dicke (BD)
theory, do not necessarily require such a variation, but provide a
natural and self-consistent framework for this variation by assuming
the existence of additional dimensions. The time variation of $G$ in
these multi-dimensional theories has recently been studied, and
their consistency with variable observational data for distant
supernova has been analyzed [8]. The variation of the gravitational
constant is found to make distant supernova appear brighter. The
recent results of Shapiro et al. [9], based on an analysis of radar
echo time delays, have set an experimental upper limit on the
possible time variation of the gravitational constant as
$|\dot{G}/G|<3\times 10^{-10} yr^{-1}$, where the dot represents a
derivative with respect to time. Theoretical calculations by Dicke
for zero-pressure Friedmann-type cosmologies, according to the BD
theory, yield $|\dot{G}/G| \approx 10^{-11} yr^{-1}$ for a flat
spacetime and $|\dot{G}/G|\approx 3\times 10^{-11} yr^{-1}$ for
closed spacetimes [10,11]. Although the time variation in $G$ is
extremely small at the present epoch, Dicke has shown that there
exist early epoch solutions for which the energy density associated
with the time variation is much greater than the matter energy
density. Recently, a constraint on the variation of $G$ has been
obtained by using the Wilkinson Microwave Anisotropy Probe (WMAP)
and the big bang nucleosynthesis observations, which comes out to be
$-3\times 10^{-13} yr^{-1}<(\dot{G}/G)_{today}<4\times 10^{-13}
yr^{-1}$ [12].

Bekenstein [13] and Bertolami [14] introduced models in which both
$G$ and $\Lambda$ are time dependent. Several authors [15-17]
studied the variations of $G$ and $\Lambda$ in the framework of flat
Friedmann-Robertson-Walker (FRW) symmetries. This work has been
extended [18-21] to more complicated geometries like the Bianchi
type I model, which is the simplest generalization of the FRW flat
model, by using a perfect fluid. The same model was considered with
viscous fluids [22-24] to discuss the time variations of $\Lambda$
and $G$. Kalligas et al. [25] discussed the behaviors of these
varying constants by using Lie method. Darabi [26] found that the
time variations of these constants lead the vacuum energy density to
be time dependent as $\rho_v=\Lambda(t)/8\pi G(t)$. Therefore, for
an early universe, where $\Lambda$ is so large and $G$ is so small
compared with their current values, the vacuum energy is huge. At
the present status of the universe, however, the vacuum energy is
vanishing due to the time variations of both $\Lambda$ and $G$.
Belinchon and Dávila [27] discussed time-varying constants in
different spacetimes by using different symmetric techniques. In
recent papers [28, 29], the same author analyzed the behaviors of
time-varying $G$ and $\Lambda$ for a Bianchi type I model and made
comparison of different techniques, including self-similarity,
matter collineations, kinematic self-similar similarity, and the Lie
method.

In this paper, we extend Belinchon's work to Bianchi type III and
Kantowski-Sachs spacetimes. We shall use two symmetric techniques,
i.e.,  matter collineations (using energy-momentum tensor for a
perfect fluid) and kinematic self-similarity (first and second
kinds). The scheme of this paper is as follows: In Section
\textbf{II}, we shall write the field equations with relevant
quantities for Bianchi type III and Kantowski-Sachs spacetimes.
Section \textbf{III} is devoted to a study of the behaviors of the
time-varying constants by using kinematic self-similarity technique.
In Section \textbf{IV}, we use the matter collineations technique to
investigate the behaviors of $G$ and $\Lambda$. The last section
will provide a summary and a discussion of the results obtained.\\

\section*{II. BIANCHI TYPE III AND KANTOWSKI-SACHS SPACETIMES}

Bianchi type III and Kantowski-Sachs spacetimes are spatially
homogeneous spacetimes that admit an abelian group of isometries
$G_3$ acting on a spacelike hypersurface. These are generated by
spacelike Killing vectors $\xi_1=\partial_r,~\xi_2=\partial_\theta$
and $\xi_3=\partial_\phi$. In co-moving coordinates, the metric
representing these spacetimes is written as [30]
\begin{equation}\label{2}
ds^2 =c^2dt^2-A^2(t)dr^2-B^2(t)(d\theta^2+f^2(\theta)d\phi^2),
\end{equation}
where $A$ and $B$ are arbitrary functions of $t$ while $f(\theta)$
is defined as
\begin{eqnarray*}
f(\theta)=\sinh\theta \quad \textmd{corresponding to Bianchi type
III spacetime,}\\\nonumber f(\theta)=\sin\theta \quad
\textmd{corresponding to Kantowski-Sachs spacetime.}
\end{eqnarray*}
These metrics represent anisotropic generalizations of the open and
closed FRW models, respectively.

The energy-momentum tensor for a perfect fluid is given by
\begin{equation}\label{3}
T_{ab}=(\rho+p)u_au_b-p g_{ab},
\end{equation}
where $u^a$ is the four-velocity and in co-moving coordinates, it is
defined as follows:
\begin{equation}\label{4}
u^a=\left(\frac{1}{c},0,0,0\right)~~ \textmd{with}~~ u^au_a=1.
\end{equation}
A perfect fluid can be characterized by a dimensionless number $k$
given by
\begin{equation}\label{5}
k=\frac{p}{\rho}\quad \textrm{or}\quad p={k}{\rho}.
\end{equation}
This is called the equation of state and represents a dust fluid for
$k=0$, radiation for $k=\frac{1}{3}$, and stiff matter for $k=1$.
Using Eq. (\ref{5}), the EFEs of Eq. (\ref{1}) lead to the following
three equations:
\begin{eqnarray}
\frac{2\dot{A}\dot{B}}{AB}-\frac{c^2f''}{B^2f
}+\frac{\dot{B}^2}{B^2}&=&\frac{8\pi
G}{c^2}\rho+\Lambda c^2,\label{6}\\
\frac{2\ddot{B}}{B}-\frac{c^2f''}{B^2f}+\frac{\dot{B}^2}{B^2}&=&-\frac{8\pi
G}{c^2}k\rho+\Lambda c^2,\label{7}\\
\frac{\ddot{A}}{A}+\frac{\ddot{B}}{B}+\frac{\dot{A}\dot{B}}{AB}&=&-\frac{8\pi
G}{c^2}k\rho+\Lambda c^2.\label{8}
\end{eqnarray}
Here, the dot denotes the time derivative, and the prime denotes the
derivative with respect to $\theta$.

The time derivatives of $G$ and $\Lambda$ can be related by the
Bianchi identities as follows:
\begin{equation}\label{9}
(R_{ab}-\frac{1}{2}Rg_{ab})^{;b}=(\frac{8\pi G}{c^4}T_{ab}+\Lambda
g_{ab})^{;b}.
\end{equation}
Simplification of this expression by fixing $a=0$ and varying
$b=0,1,2,3$ yields
\begin{equation}\label{10}
\frac{8\pi}{c^4}\dot{G}\rho+\dot{\Lambda}=\frac{8{\pi}G}{c^4}
\left[\dot{\rho}+\rho(1+k)\left(\frac{\dot{A}}{A}+
2\frac{\dot{B}}{B}\right)\right].
\end{equation}
The conservation law of the energy-momentum tensor of matter field,
$T^{ab}_{;b}=0$, gives
\begin{equation}\label{11}
\dot{\rho}+\rho(1+k)\left(\frac{\dot{A}}{A}+2\frac{\dot{B}}{B}\right)=0.
\end{equation}
In view of this equation, Eq. (\ref{10}) implies that
\begin{equation}\label{12}
\dot{\Lambda}=-\frac{8\pi}{c^4}\dot{G}\rho.
\end{equation}
It is mentioned here that for the possibilities $a=1, 2, 3$ and
$b=0, 1, 2, 3$, Eq. (\ref{9}) is satisfied identically.

We define Hubble's parameter as an average expansion of the universe
as follows:
\begin{equation}\label{13}
H=\frac{\dot{A}}{A}+2\frac{\dot{B}}{B}=H_1+2H_2,
\end{equation}
where $$H_1=\frac{\dot{A}}{A},\quad H_2=\frac{\dot{B}}{B}.$$
Consequently, the deceleration parameter is defined as
\begin{equation}
q=\frac{d}{dt}(\frac{1}{H})-1.
\end{equation}
Using Eq. (\ref{13}) in Eq. (\ref{11}), it follows that
\begin{equation}\label{15}
\dot{\rho}+\rho(1+k)H=0.
\end{equation}
In the next two sections, we use a Bianchi type III spacetime to
discuss the behaviors of $G$ and $\Lambda$. However, we include the
discussion of the behaviors of $G$ and $\Lambda$ for Kantowski-Sachs
spacetime in the last section.

\begin{center}
\section*{III. TIME-VARYING BEHAVIOR OF $G$ AND $\Lambda$ BY USING
A KINEMATIC SELF-SIMILARITY TECHNIQUE}
\end{center}

In general relativity, self-similarity can be defined by the
existence of a homothetic vector field. Cahill and Taub [31] were
the pioneers to introduce the concept of self-similarity
corresponding to homothety. Carter and Henriksen [32] introduced the
concept of kinematic self-similarity as a natural generalization of
the homothetic case. A kinematic self-similar vector field $\xi$
satisfies the following conditions [33]:
\begin{eqnarray}
\pounds_\xi u_a =\alpha u_a,\label{16}\\
\pounds_\xi h_{ab} = 2\delta h_{ab},\label{17}
\end{eqnarray}
where $\alpha$ and $\delta$ are constants and $h_{ab}=g_{ab}-u_au_b$
is the projection tensor. Kinematic self-similarity (KSS) can be
classified into first and second kinds by using a scale independent
ratio $\frac{\alpha}{\delta}$, referred to as the \emph{similarity
index}. The ratio $\frac{\alpha}{\delta}=1$ leads to self-similarity
of the first kind, also known as homothety, while the ratio
$\frac{\alpha}{\delta}\neq 0, 1$ indicates self-similarity of the
second kind. Here, we discuss the behaviors of $G$ and $\Lambda$ by
using these kinds.\\

\subsection*{1. Kinematic Self-similaity of the First Kind}

For $\alpha=\delta=1$, the KSS vector field yields
\begin{equation}\label{18}
g_{ab,c}\xi^c+ g_{ac}\xi^c_{,b}+ g_{bc}\xi^c_{,a}=2g_{ab}.
\end{equation}
In this case, $\xi$ is known as a homothetic vector field. This
gives the following system of ten equations:
\begin{eqnarray}\label{E'00}
\xi^0_{,0}&=&1,\\\label{E'01}
c^2\xi^0_{,1}-A^2\xi^1_{,0}&=&0,\\\label{E'02}
c^2\xi^0_{,2}-B^2\xi^2_{,0}&=&0,\\\label{E'03}
c^2\xi^0_{,3}-B^2\sinh^2\theta \xi^3_{,0}&=&0,\\\label{E'11}
(\frac{\dot{A}}{A})\xi^0+\xi^1_{,1}&=&1,\\\label{E'12}
A^2\xi^1_{,2}+B^2\xi^2_{,1}&=&0,\\\label{E'13}
A^2\xi^1_{,3}+B^2\sinh^2\theta \xi^3_{,1}&=&0,\\\label{E'22}
(\frac{\dot{B}}{B})\xi^0+\xi^2_{,2}&=&1,\\\label{E'23}
B^2\xi^2_{,3}+B^2\sinh^2\theta \xi^3_{,2}&=&0,\\\label{E'33}
(\frac{\dot{B}}{B})\xi^0+\xi^3_{,3}&=&1.
\end{eqnarray}
Here, the derivatives with respect to 0, 1, 2, and 3 denote the
partial derivatives with respect to $t,r,\theta$, and $\phi$,
respectively. Solving Eqs. (\ref{E'00}), (\ref{E'11}), (\ref{E'22}),
and (\ref{E'33}) simultaneously, we obtain
\begin{eqnarray}
\xi^0&=&t+f_0(r,\theta,\phi),\label{xi0}\\
\xi^1&=&(1-\frac{\dot{A}}{A}\xi^0)r+f_1(t,\theta,\phi),\label{xi1}\\
\xi^2&=&(1-\frac{\dot{B}}{B}\xi^0)\theta+f_2(t,r,\phi),\label{xi2}\\
\xi^3&=&(1-\frac{\dot{B}}{B}\xi^0)\phi+f_3(t,r,\theta),\label{xi3}
\end{eqnarray}
where $f_1(t,\theta,\phi)-f_3(t,r,\theta)$ are functions of
integration. Substituting these values of $\xi$ in Eqs. (\ref{E'11})
and (\ref{E'22}), respectively, we obtain
\begin{eqnarray}
2A\dot{A}f_{0,1}(r,\theta,\phi)r=0,\label{33}\\
2B\dot{B}f_{0,2}(r,\theta,\phi)\theta=0.\label{34}
\end{eqnarray}
Since $A,B,r,$ and $\theta \neq0$, the above equations yield the
following four cases:\\
$$(i)~\dot{A}=0=\dot{B},~(ii)~\dot{A}=0,~\dot{B}\neq0,~
(iii)~\dot{A}\neq0,~ \dot{B}=0,~(iv)~\dot{A },~\dot{B}\neq0.$$ The
cases (i)-(iii) do not provide the behaviors of $G$ and $\Lambda$ as
they vanish. In the following, we discuss the case when
$\dot{A},~\dot{B}\neq0$.

In this case, Eqs. (\ref{33}) and (\ref{34}) yield
$$f_0(r,\theta,\phi) \equiv f_0(\phi),$$
which, in view of Eq. (\ref{E'33}), finally gives
$$f_0(\phi)\equiv c_0,$$
where $c_0$ is an arbitrary constant. Thus, $\xi^0$ takes the form
\begin{equation}
\xi^0=t+c_0.\label{37}
\end{equation}
When we make use of this value of $\xi^0$ in Eqs.
(\ref{xi1})-(\ref{xi3}), it follows that
\begin{eqnarray}
\xi^1&=&(1-\frac{\dot{A}}{A}(t+c_0))r+f_1(t,\theta,\phi),\label{38}\\
\xi^2&=&(1-\frac{\dot{B}}{B}(t+c_0))\theta+f_2(t,r,\phi),\label{39}\\
\xi^3&=&(1-\frac{\dot{B}}{B}(t+c_0))\phi+f_3(t,r,\theta).\label{40}
\end{eqnarray}

Now replacing these components of $\xi$ in the above system of
equations, the functions of integration
$f_1(t,\theta,\phi)-f_3(t,r,\theta)$ reduce to arbitrary constants
$c_1,~ c_2,$ and $c_3$ respectively along with the following ODEs as
necessary and sufficient conditions:
\begin{eqnarray}
\dot{A}A+(t+c_0)(\ddot{A}A-\dot{A}^2)=0,\label{41}\\
\dot{B}B+(t+c_0)(\ddot{B}B-\dot{B}^2)=0.\label{42}
\end{eqnarray}
Thus, the homothetic vector field becomes
\begin{eqnarray}\label{43}
\xi&=&(t+c_0)\partial_t+[(1-(t+c_0)\frac{\dot{A}}{A})r+c_1]\partial_r
+[(1-(t+c_0)\frac{\dot{B}}{B})\theta+c_2]\partial_\theta\nonumber\\
&+&[(1-(t+c_0)\frac{\dot{B}}{B})\phi+c_3]\partial_\phi.
\end{eqnarray}
Using Eq. (\ref{13}), the solution of the ODEs in Eqs. (\ref{41})
and (\ref{42}), respectively, yield
\begin{equation}\label{44}
A=A_0(t+c_0)^{\alpha_1},\quad B=B_0(t+c_0)^{\alpha_2},
\end{equation}
where $A_0,~B_0,~\alpha_1$, and $\alpha_2$ are positive constants
(for physical reasons) of integration. Making use of these values of
$A$ and $B$ in Eq. (\ref{13}), the Hubble and the deceleration
parameters turn out to be
\begin{eqnarray}\label{45}
H&=&(\alpha_1+2\alpha_2)(t+c_0)^{-1},\\
q&=&\frac{1}{\alpha_1+2\alpha_2}-1.
\end{eqnarray}
Consequently, Eq. (\ref{15}) yields
\begin{equation}\label{46}
\rho=\rho_0(t+c_0)^{-(1+k)(\alpha_1+2\alpha_2)},
\end{equation}
where $\rho_0$ is a constant of integration.

Solving Eqs. (\ref{6}) and (\ref{7}) simultaneously and then using
Eqs. (\ref{44}) and (\ref{46}), we have
\begin{eqnarray}
G&=&\frac{2c^2}{8\pi\rho_0(1+k)}
(\alpha_1\alpha_2-\alpha_2(\alpha_2-1))
(t+c_0)^{-2+(\alpha_1+2\alpha_2)(1+k)},\label{47}\\
\Lambda &=&\frac{1}{c^2}[2\alpha_1\alpha_2+{\alpha_2}^2-\frac{2}
{1+k}(\alpha_1\alpha_2-{\alpha_2}^2+\alpha_2)](t+c_0)^{-2}\nonumber\\
&-&\frac{1}{B_0^2}(t+c_0)^{-2\alpha_2}.\label{48}
\end{eqnarray}
Using the values of $G,$ and $\Lambda$ and Eq. (\ref{46}) in Eq.
(\ref{8}), it follows that
$$\alpha_2(\alpha_2-1)+\alpha_1\alpha_2+\alpha_1(\alpha_1-1)=
2\alpha_2(\alpha_2-1)+{\alpha_2}^2-\frac{c^2}{{B_0}^2}(t+c_0)^{-2\alpha_2+2}.$$
This shows that $\alpha_1$ will be constant only if we choose
$\alpha_2=1$. Thus, we have
\begin{equation}\label{50}
\alpha_1=\sqrt{1-\frac{c^2}{{B_0}^2}}.
\end{equation}
For $\alpha_1$ to be real, it is necessary that
$${B_0}^2 \geq c^2\quad\Rightarrow\quad B_0\geq c \quad \textmd{or}
\quad B_0\leq-c.$$ In other words, we can say that
\begin{equation}\label{52}
B_0 \in (-\infty,-c]\cup[c,\infty).
\end{equation}
For $B_0$ to be positive, our interval of interest is only where
$B_0 \in [c,\infty).$

\subsubsection*{i. Behavior of $G$}

Using $\alpha_2=1$ in Eq. (\ref{47}), $G$ becomes
\begin{equation}\label{53}
G=G_0(t+c_0)^{\alpha_1+(\alpha_1+2)k},
\end{equation}
where
\begin{equation}\label{54}
G_0=\frac{2c^2\alpha_1}{8\pi\rho_0(1+k)}.
\end{equation}
We note that Eqs. (\ref{46}) and (\ref{53}) yield
\begin{equation}\label{55}
G \rho \approx (t+c_0)^{-2}.
\end{equation}

For $G_0>0$, i.e., $\alpha_1=\sqrt{1-\frac{c^2}{{B_0}^2}}$, the
behavior of $G$ can be discussed as follows:
\begin{eqnarray}
G ~\textmd{is}~ \textmd{increasing}~&\Leftrightarrow & ~
\alpha_1>\frac{-2k}{(1+k)} \nonumber\\
&\Leftrightarrow & \quad
\sqrt{1-\frac{c^2}{{B_0}^2}}>\frac{-2k}{1+k}.\label{2.3.12}
\end{eqnarray}
When we take square of both sides, the above inequality yields
the following two cases:\\
\\
\textbf{(i)} Here, the inequality in Eq. (\ref{2.3.12}) implies that
$$1-\frac{c^2}{{B_0}^2}> \frac{4k^2}{(1+k)^2} \quad \textmd{if}
\quad \sqrt{1-\frac{c^2}{{B_0}^2}}> \frac{2k}{(1+k)}.$$ Thus,
\begin{eqnarray}
G ~\textmd{is}~ \textmd{increasing}~&\Leftrightarrow& \quad
{B_0}^2>{\left(\frac{c(1+k)}{\sqrt{1+2k-3k^2}}\right)}^2\nonumber\\
& \Leftrightarrow & \quad B_0\in(-\infty,-b)\cup(b,\infty),
\end{eqnarray}
where
\begin{equation}
b=\frac{c(1+k)}{\sqrt{1+2k-3k^2}}, \quad \forall \quad k \in
(-\frac{1}{3}, 1); \label{2.2.121a}
\end{equation}
$b$ becomes imaginary or infinite for all other values of $k$. One
can easily verify that $b>c \quad \forall \quad k \in
(-\frac{1}{3}, 1)$.\\
\\
\textbf{(ii)} The inequality in Eq. (\ref{2.3.12}) implies that
$$
1-\frac{c^2}{{B_0}^2}< \frac{4k^2}{(1+k)^2} \quad \textmd{if} \quad
\sqrt{1-\frac{c^2}{{B_0}^2}}< \frac{2k}{(1+k)}.$$ Thus,
\begin{eqnarray}
G ~\textmd{is}~ \textmd{increasing}~&\Leftrightarrow& \quad
{B_0}^2< {\left(\frac{c(1+k)}{\sqrt{1+2k-3k^2}}\right)}^2 \nonumber\\
& \Leftrightarrow & \quad B_0\in(-b,-c)\cup(c,b), \label{2.2.121}
\end{eqnarray}
where $b$ is the same as defined in Eq. (\ref{2.2.121a}). Hence,
from both the cases, we can conclude that $G$ increases for all
$B_0\in \Re^+ \backslash (0,c)$ while $G$ becomes constant at $ b$.
It is mentioned here that $G$ always vanishes at $B_0=c$.

Now, we discuss the behaviors of $G$ in the dust, radiation, and
stiff fluid cases. For dust, we take $k=0$, and the behavior of $G$
is the following:
\begin{eqnarray}
G \quad\textmd{ is}\quad \textmd{increasing}~
&\Leftrightarrow& ~ B_0\in \Re^+ \backslash (0,c),\nonumber\\
G\quad \quad \textmd{vanishes}\quad&\Leftrightarrow&~
B_0=c.\nonumber
\end{eqnarray}
For the radiation case, we have $k=1/3$; thus,
\begin{eqnarray}
G \quad \textmd{is}\quad \textmd{increasing}~ &\Leftrightarrow& ~
B_0\in \Re^+ \backslash (0,\frac{2c}{\sqrt{3}}),\nonumber\\
G \quad\textmd{is}\quad\textmd{constant}~&\Leftrightarrow&
~B_0=\frac{2c}{\sqrt{3}}.\nonumber
\end{eqnarray}
For the stiff fluid, $G$ increases for all values of $B_0 \in \Re^+
\backslash (0,c)$.

\subsubsection*{ii. Behavior of $\Lambda$}

To discuss the behavior of $\Lambda$, we substitute $\alpha_2=1$ in
Eq. (\ref{48}) so that $\Lambda$ becomes
\begin{equation}\label{65}
\Lambda=\left[\frac{(2\alpha_1+1)(1+k){B_0}^2-c^2(1+k)
-2\alpha_1{B_0}^2}{c^2{B_0}^2(1+k)}\right](t+c_0)^{-2}=\Lambda_0(t+c_0)^{-2},
\end{equation}
where
\begin{equation}\label{66}
\Lambda_0=\frac{(2\alpha_1+1)(1+k){B_0}^2-c^2(1+k)-2\alpha_1{B_0}^2}{c^2{B_0}^2(1+k)}
\end{equation}
is a constant. From this equation, we can discuss the behavior of
$\Lambda$ as follows:
\begin{eqnarray}\label{68}
\Lambda~~\textmd{is~~increasing} ~~ &\textmd{if}&\quad
\Lambda_0<0 \quad \textmd{and}\quad t>c_0,\nonumber\\
&\textmd{or}&\quad \Lambda_0>0 \quad \textmd{and}\quad t<c_0,\nonumber\\
\Lambda~~\textmd{is~~decreasing}~~&\textmd{if}&\quad
\Lambda_0>0 \quad \textmd{and} \quad t>c_0,\nonumber\\
&\textmd{or}&\quad \Lambda_0<0 \quad \textmd{and}\quad t<c_0,\nonumber\\
\Lambda\quad~\textmd{vanishes}\quad~~&\textmd{if}&\quad
\Lambda_0=0~~\textmd{or}~~ t\rightarrow\infty.
\end{eqnarray}
It is noticed that $\Lambda_0>0$ if
$$(1+k)(2\alpha_1{B_0}^2+({B_0}^2-c^2))>2\alpha_1{B_0}^2.$$
Using Eq. (\ref{50}), this inequality becomes
\begin{eqnarray}\label{2.2.130}
(1+k)(2\alpha_1+\alpha_1^2)>2\alpha_1,\\\nonumber
\Rightarrow~~~\alpha_1>\frac{-2k}{1+k},
\end{eqnarray}
 which is the same condition as given by the inequality
in Eq. (\ref{2.3.12}). Hence, we finally obtain
\begin{equation}
\Lambda_0>0 \quad \forall\quad B_0 \in \Re^+ \backslash (0,c).
\end{equation}
Similarly, $\Lambda_0<0$ if
$$(1+k)(2\alpha_1+\alpha_1^2)<2\alpha_1,$$
which yields a contradiction. Further, $\Lambda_0$ vanishes at
$B_0=b$.

For different types of fluids, we can discuss the above conditions
on $\Lambda_0$ as follows: In the dust and stiff fluid case,
$\Lambda_0$ is always positive for any value of $B_0 \in \Re^+
\backslash (0,c)$ while in the radiation case $\Lambda_0$ is
positive for all $B_0 \in {\Re}^+ \backslash
(0,\frac{2c}{\sqrt{3}}).$

\subsection*{2. Kinematic Self-similarity of the Second Kind}

For $\alpha=\delta \neq 0,1$, the definition of KSS vector field
yields
\begin{eqnarray}\label{S'00}
\xi^0_{,0}&=&2\alpha c^2,\\\label{S'01}
c^2\xi^0_{,1}-A^2\xi^1_{,0}&=&0,\\\label{S'02}
c^2\xi^0_{,2}-B^2\xi^2_{,0}&=&0,\\\label{S'03}
c^2\xi^0_{,3}-B^2\sinh^2\theta\xi^3_{,0}&=&0,\\\label{S'11}
(\frac{\dot{A}}{A})\xi^0+\xi^1_{,1}&=&2\delta,\\\label{S'12}
A^2\xi^1_{,2}+B^2\xi^2_{,1}&=&0,\\\label{S'13}
A^2\xi^1_{,3}+B^2\sin^2\theta\xi^3_{,1}&=&0,\\\label{S'22}
(\frac{\dot{B}}{B})\xi^0+\xi^2_{,2}&=&2\delta ,\\\label{S'23}
\xi^2_{,3}+\sinh^2\theta\xi^3_{,2}&=&0,\\\label{S'33}
(\frac{\dot{B}}{B})\xi^0+\xi^3_{,3}&=&2\delta.
\end{eqnarray}
Solving this system of equations simultaneously by adopting the same
procedure as in the first kind of KSS, it follows that
\begin{eqnarray}\label{85}
\xi&=&(\alpha t+\beta)\partial_t+[(\delta-\frac{\dot{A}}{A}(\alpha
t+\beta))r+c_5]\partial_r\nonumber\\
&+&[(\delta-\frac{\dot{B}}{B}(\alpha
t+\beta))\theta+c_6]\partial_\theta+[(\delta-\frac{\dot{A}}{A}(\alpha
t+\beta))\phi+c_7]\partial_\phi.\\
A&=&A_0(t+\frac{\beta}{\alpha})^{\alpha_1},~~B=B_0(t+\frac{\beta}{\alpha})^{\alpha_2},\label{86}\\
H&=&(\alpha_1+2\alpha_2)(t+\frac{\beta}{\alpha})^{-1},\label{88}\\
\rho&=&\rho_0(t+\frac{\beta}{\alpha})^{-(1+k)(\alpha_1+2\alpha_2)},\label{89}
\end{eqnarray}
where $A_0$, $B_0$, $\alpha_1$, $\alpha_2$, and $\rho_0$ are
constants of integration.

Similarly, $G$ and $\Lambda$ can be found to be
\begin{eqnarray}
G&=&G_0(t+\frac{\beta}{\alpha})^{\alpha_1+(\alpha_1+2)k},\label{100}\\
\Lambda&=&\Lambda_0(t+\frac{\beta}{\alpha})^{-2},\label{101}
\end{eqnarray}
where $G_0$ and $\Lambda_0$ are the same as given in the first kind.
Obviously, Eqs. (\ref{100}) and (\ref{101}) reduce exactly to Eqs.
(\ref{53}) and (\ref{65}) for $c_0=\frac{\beta}{\alpha}$; hence, the
behavior of $G$ and $\Lambda$ will be the same as in case of KSS of
the first kind just by replacing $c_0=\frac{\beta}{\alpha}$ in the
corresponding constraint equations.

\begin{center}
\section*{IV. TIME-VARYING BEHAVIOR OF $G$ and $\Lambda$ BY USING THE
MATTER COLLINEATION TECHNIQUE}
\end{center}
The energy-momentum tensor
$T_{ab}$ represents the matter part of the field equations
in Eq. (\ref{1}). This enables us to understand the physical structure of
spacetime. Symmetries of the energy-momentum tensor provide the
conservation laws on matter fields. Matter collineation is defined
as
\begin{equation}\label{102}
\pounds_\xi T_{ab}=0.
\end{equation}
This does not give any information about the behaviors of $G$ and
$\Lambda$. However, it can be modified to obtain the behavior of $G$
as follows [28]:
\begin{equation}\label{103}
\pounds_\xi(\frac{8\pi G(t)}{c^4}T_{ab})=0.
\end{equation}
If we introduce $\Lambda$, we get a complete modification of matter
collineation, Eq. (\ref{102}) [28]:
\begin{equation}\label{104}
\pounds_\xi(\frac{8\pi G(t)}{c^4}T_{ab}+{\Lambda(t)}g_{ab})=0.
\end{equation}
This equation helps us to discuss the behaviors of time-dependent
$G$ and $\Lambda$.\\

\subsection*{1. Matter Collineations}

Using the components of a perfect fluid energy-momentum tensor, the
system of matter collineation (MC) equations yields
\begin{eqnarray}
(\rho c^2)_{,0}\xi^0+2\rho c^2\xi^0_{,0}&=&0,\label{M'00}\\
\rho c^2\xi^0_{,1}+pA^2\xi^1_{,0}&=&0,\label{M'01}\\
\rho c^2\xi^0_{,2}+pB^2\xi^2_{,0}&=&0,\label{M'02}\\
\rho c^2\xi^0_{,3}+pB^2\sinh^2\theta \xi^3_{,0}&=&0,\label{M'03}\\
(pA^2)_{,0}\xi^0+2pA^2\xi^1_{,1}&=&0,\label{M'11}\\
pA^2\xi^1_{,2}+pB^2\xi^2_{,1}&=&0,\label{M'12}\\
pA^2\xi^1_{,3}+pB^2\sinh^2\theta \xi^3_{,1}&=&0,\label{M'13}\\
(pB^2)_{,0}\xi^0+2pB^2\xi^2_{,2}&=&0,\label{M'22}\\
pB^2\xi^2_{,3}+pB^2\sinh^2\theta \xi^3_{,2}&=&0,\label{M'23}\\
(pB^2\sinh^2\theta)_{,0}\xi^0+2pB^2\sinh^2\theta
\xi^3_{,3}&=&0.\label{M'33}
\end{eqnarray}
It follows from Eq. (\ref{M'00}) that
\begin{equation}\label{115}
\xi^0=f_8(r,\theta,\phi)\rho^{-1/2},
\end{equation}
where $f_8(r,\theta,\phi)$ is a function of integration. Similarly,
Eq. (\ref{M'11}) yields
$$(\frac{\dot{p}}{p}+2\frac{\dot{A}}{A})\xi^0+2\xi^1_{,1}=0.$$
Using Eqs. ({\ref{5}) and (\ref{115}), this equation gives
\begin{equation}\label{118}
\xi^1=\xi^0\left(\frac{{\dot{\xi}}^0}{\xi^0}
-\frac{\dot{A}}{A}\right)r+f_9(t,\theta,\phi),
\end{equation}
where $f_9(t,\theta,\phi)$ is an integration function. Similarly, by
solving Eqs. (\ref{M'22}) and (\ref{M'33}) and then integrating
w.r.t. $\theta$ and $\phi$, respectively, we obtain
\begin{eqnarray}\label{119}
\xi^2&=&\xi^0\left(\frac{{\dot{\xi}}^0}{\xi^0}
-\frac{\dot{B}}{B}\right)\theta+f_{10}(t,r,\phi),\\
\xi^3&=&\xi^0\left(\frac{{\dot{\xi}}^0}{\xi^0}
-\frac{\dot{B}}{B}\right)\phi+f_{11}(t,r,\theta),\label{120}
\end{eqnarray}
where $f_{10}(t,r,\phi)$ and $f_{11}(t,r,\theta)$ are functions of
integration.

When we solve the above system of ten equations simultaneously by
using these values of $\xi$, as done in the KSS technique, the
functions of integration, $f_8(r,\theta,\phi)-f_{11}(t,r,\theta)$,
reduce to arbitrary constants, which are termed as $c_8, c_9,
c_{10},$ and $c_{11}$, respectively. Thus, the final form of the
vector field, in terms of $\xi^0=c_8\rho^{-1/2}$, can be written as
\begin{equation}\label{121}
\xi=\xi^0\partial_t+[({\dot{\xi}}^0-\frac{\dot{A}}{A}\xi^0)r+c_9]\partial_r+
[({\dot{\xi}}^0-\frac{\dot{B}}{B}{\xi^0})\theta+c_{10}]\partial_\theta+
[({\dot{\xi}}^0-\frac{\dot{B}}{B})\phi+c_{11}]\partial_\phi,
\end{equation}
along with the following constraint equations
\begin{eqnarray}\label{122}
p\partial_t[\xi^0(\frac{\dot{\rho}}{2\rho}+\frac{\dot{A}}{A})]=0,\quad
p\partial_t[\xi^0(\frac{\dot{\rho}}{2\rho}+\frac{\dot{B}}{B})]=0.
\end{eqnarray}
These two equations reveal that either $p=0$, i.e., a dust case, or
$p\neq 0$, i.e., a perfect fluid case. The dust case is not
interesting as the time-varying behaviors, of $G$ and $\Lambda$
cannot be discussed for this case by using MCs.

In the perfect fluid case, Eq. (\ref{122}) implies that
\begin{equation}
A=\frac{1}{\sqrt{\rho}}e^{a_1\int{\sqrt{\rho}dt}+
a_2},~~~B=\frac{1}{\sqrt{\rho}}e^{b_1\int{\sqrt{\rho}dt}+b_2},\label{125}
\end{equation}
where $a_1, a_1, b_1$, and $b_2$ are arbitrary constants of
integration. For these values of the metric functions, the EFEs give
\begin{eqnarray}\label{126}
G&=&\frac{2c^2}{8\pi\rho(1+k)}[\frac{\ddot{\rho}}{\rho^2}
-\frac{{\dot{\rho}}^2}{\rho^3}+c_{13}\frac{\dot{\rho}}{\rho^{3/2}}
-\frac{b_1}{\rho^{3/2}}+2c_{14}],\\
\Lambda&=&\frac{1}{c^2}[(b_1\rho^{1/2}-\frac{\dot{\rho}}{2\rho})^2
-(\frac{c}{{\rho}^{1/2}}e^{-b_1\int{\sqrt{\rho}dt}+b_2})^2\nonumber\\
&+&\frac{2k}{1+k}(b_2\rho-c_{15}\frac{\dot{\rho}\rho^{1/2}}{2}
+\frac{\dot{\rho}^2\rho^{-2}}{4})\nonumber\\
&+&\frac{2}{1+k}(c_{13}\rho+\frac{3\dot{\rho}^2}{4\rho^2}-b_1
(\frac{\dot{\rho}}{\rho^{1/2}}-\frac{1}{2\rho^{-1/2}})
-\frac{\ddot{\rho}}{\rho}].\label{127}
\end{eqnarray}
These equations show that the time-varying behaviors of $G$ and
$\Lambda$ cannot be discussed unless $\rho$ is given. We assume the
following two cases to discuss the behaviors of $G$ and
$\Lambda$:\\
\\
\textbf{(i)} $\rho(t)=\rho_0 ~(\textmd{constant})$;\\
\textbf{(ii)} $\rho(t)=\frac{1}{(at+b)^2}$, where
$a=\frac{1}{c_8}$ and $b=\frac{c_0}{c_8}$.\\\\
\textbf{Case (i)} When we substitute $\rho(t)=\rho_0$ (constant) in
Eqs. (\ref{126}) and (\ref{127}), it follows that
\begin{equation}\label{128}
G=\frac{2c^2}{8\pi\rho_0(1+k)}\left(\frac{-b_1}{\rho_0^{3/2}}+2c_{18}\right)
\end{equation}
and
\begin{equation}\label{129}
\Lambda=\frac{-1}{\rho_0}e^{-2b_1(t+b_2)}+\Lambda_0,
\end{equation}
where
$$\Lambda_0=\frac{1}{c^2}[(b_1\rho_0^{1/2})^2+\frac{2k}{1+k}b_2\rho_0
+\frac{2}{1+k}(c_{13}\rho_0+\frac{b_1}{2\rho_0^{-1/2}})].$$
We see from Eqs. (\ref{128}) and (\ref{129}) that $G$ always remains
constant while the behavior of $\Lambda$ is as follows:
\begin{eqnarray}\label{131}
\Lambda~~\textmd{is~~increasing} ~~ &\Leftrightarrow&\quad b_1>0,\nonumber\\
\Lambda~~\textmd{is~~decreasing}~~&\Leftrightarrow&\quad
b_1<0,\nonumber\\
\Lambda~~\textmd{is~~constant}~~~~&\Leftrightarrow&\quad b_1=0.
\end{eqnarray}\\
\textbf{Case (ii)} In this case, we get $\xi^0=t+c_0$. Using this in
Eq. (\ref{121}), we obtain the same homothetic vector field as in
case of KSS of the first kind given in Eq. (\ref{43}). It is
mentioned here that this homothetic vector field satisfies the relation
\begin{equation}\label{132}
\pounds_{HO}(T_{ab})=0
\end{equation}
with the same ODEs as given by Eqs. (\ref{41}) and (\ref{65}). Thus,
the behaviors of $G$ and $\Lambda$ will be the same as in the case
of KSS of the first kind.

\subsection*{2. Modified Matter Collineations}

The modified MC equations, Eq. (\ref{103}), can be written as
\begin{equation}\label{133}
\xi^o\dot{G}T_{ab}+G(T_{ab,c}\xi^c
+T_{ac}\xi^c_{,b}+T_{bc}\xi^c_{,a})=0.
\end{equation}
The corresponding system of MC equations yields
\begin{eqnarray}
\xi^0&=&c_{17}(G\rho)^{-1/2},\label{134}\\
\xi^1&=&\xi^0\left(\frac{{\dot{\xi}}^0}{\xi^0}
-\frac{\dot{A}}{A}\right)r+c_{18},\label{135}\\
\xi^2&=&\xi^0\left(\frac{{\dot{\xi}}^0}{\xi^0}
-\frac{\dot{B}}{B}\right)\theta+c_{19},\label{136}\\
\xi^3&=&\xi^0\left(\frac{{\dot{\xi}}^0}{\xi^0}
-\frac{\dot{B}}{B}\right)\phi+c_{20}.\label{137}
\end{eqnarray}
Here, $c_{17},c_{18},c_{19}$, and $c_{20}$ are arbitrary constants of
integration satisfying the following ODEs
\begin{eqnarray}\label{138}
p\partial_t[\xi^0(\frac{1}{2}(\frac{\dot{\rho}}{\rho}
+\frac{\dot{G}}{G})+\frac{\dot{A}}{A})]&=&0,\\
p\partial_t[\xi^0(\frac{1}{2}(\frac{\dot{\rho}}
{\rho}+\frac{\dot{G}}{G})+\frac{\dot{B}}{B})]&=&0.\label{139}
\end{eqnarray}
For $p\neq0$, Eqs. (\ref{138}) and (\ref{139}) provide
\begin{equation}
A=\frac{1}{\sqrt{G \rho}}e^{c_{21}\int{\sqrt{G
\rho}~dt}+c_{22}},~~~B=\frac{1}{\sqrt{G\rho}}
e^{c_{23}\int{\sqrt{G\rho}~dt}+c_{24}}.\label{141}
\end{equation}
For these values of $A$ and $B$,  the EFEs yield
\begin{eqnarray}
G&=&\frac{2c^2}{8\pi\rho(1+k)}[\{c_{21}(G\rho)^{1/2}
-\frac{1}{2}(\frac{\dot{\rho}}{\rho}+\frac{\dot{G}}{G})\}
\{c_{25}(G\rho)^{1/2}-\frac{1}{2}(\frac{\dot{\rho}}{\rho}+\frac{\dot{G}}{G})\}\nonumber\\
&-&\{c_{25}(G\rho)^{1/2}-\frac{1}{2}(\frac{\dot{\rho}}{\rho}+\frac{\dot{G}}{G})\}^2
-\frac{1}{2}c_{25}(\dot{G}\rho+G\dot{\rho})(G\rho)^{-1/2}\nonumber\\
&-&\frac{1}{2}(\frac{\ddot{\rho}}{\rho}-\frac{\dot{\rho^2}}{\rho^2}
+\frac{\ddot{G}}{G}-\frac{\dot{G}^2}{G^2})],\label{142}\\
\Lambda&=&\frac{1}{c^2}[\{c_{25}(G\rho)^{1/2}
-\frac{1}{2}(\frac{\dot{\rho}}{\rho}+\frac{\dot{G}}{G})\}^2
-\{\frac{c}{(G\rho)^{1/2}}e^{-c_{24}\int{\sqrt{G\rho}dt}+c_{25}}\}^2\nonumber\\
&+&\frac{2k}{1+k}\{c_{21}(G\rho)^{1/2}-\frac{1}{2}(\frac{\dot{\rho}}{\rho}
+\frac{\dot{G}}{G})\}\{c_{25}(G\rho)^{1/2}
-\frac{1}{2}(\frac{\dot{\rho}}{\rho}+\frac{\dot{G}}{G})\}\nonumber\\
&+&\frac{2}{1+k}\{(c_{25}(G\rho)^{1/2}
-\frac{1}{2}(\frac{\dot{\rho}}{\rho}+\frac{\dot{G}}{G}))^2
+\frac{1}{2}c_{25}(\dot{G}\rho+G\dot{\rho})(G\rho)^{-1/2}\nonumber\\
&-&\frac{1}{2}(\frac{\ddot{\rho}}{\rho}-\frac{\dot{\rho^2}}{\rho^2}
+\frac{\ddot{G}}{G}-\frac{\dot{G}^2}{G^2})\}].\label{143}
\end{eqnarray}
It is obvious from Eqs. (\ref{142}) and (\ref{143}) that the
expressions of $G$ and $\Lambda$ are too complicated to discuss. Thus,
we assume $\xi^0=t+c_0$, i.e., $G(t) \rho(t)=\frac{1}{(at+b)^2}$ (as
found in homothetic case), such that the vector field becomes
homothetic and satisfies the equation
\begin{equation}\label{144}
\pounds_{HO}\left(\frac{8 \pi G(t)}{c^4}T_{ab}\right)=0
\end{equation}
along with the same ODEs as given by Eqs. (\ref{41}) and (\ref{42}).
This corresponds to the KSS of the first kind. It is worth mentioning
here that this case leads to
\begin{equation}\label{145}
\frac{\dot{G}}{G}+\frac{\dot{\rho}}{\rho}=\frac{-2}{(t+c_0)} ~
\Leftrightarrow ~ G \rho \approx (t+c_0)^{-2},
\end{equation}
which is exactly the same relation for the product $G\rho$ as found
by using the KSS technique.

\subsection*{3. Completely Modified Matter Collineations}

This is the case in which the variation of $\Lambda$ is included, as
well. Here, we consider Eq. (\ref{104}) and apply the definition of
the Lie derivative to obtain
\begin{eqnarray}\label{146}
&&\xi^0\dot{G}T_{ab}+G(T_{ab,c}\xi^c
+T_{ac}\xi^c_{,b}+T_{bc}\xi^c_{,a})]\nonumber\\
&&+[\dot{\Lambda}\xi^0 g_{ab}+\Lambda(g_{ab,c}\xi^c
+g_{ac}\xi^c_{,b}+g_{bc}\xi^c_{,a})=0.
\end{eqnarray}
The system of equations here leads to the following solution:
\begin{eqnarray*}
\frac{\dot{\xi^0}}{\xi}=\frac{-1}{2}\left[\frac{\frac{8\pi}{c^4}(\dot{G}\rho
+\dot{\rho}G)+\dot{\Lambda}}
{\frac{8\pi}{c^4}(G\rho)+\Lambda}\right],
\end{eqnarray*}
giving
\begin{equation}\label{148}
\xi^0=c_{26}(r,\theta,\phi)\left(\frac{8 \pi
G\rho}{c^4}+\Lambda\right)^{-1/2}.
\end{equation}
The remaining components of $\xi$ are
\begin{eqnarray}
\xi^1&=&\frac{-1}{2}\left[\frac{\frac{8\pi}{c^4}(\dot{G}p+\dot{p}G)-\dot{\Lambda}}
{\frac{8\pi}{c^4}(Gp)-\Lambda}+\frac{2\dot{A}}{A}\right]\xi^0r+c_{27}(t,\theta,\phi),\label{149}\\
\xi^2&=&\frac{-1}{2}\left[\frac{\frac{8\pi}{c^4}(\dot{G}p+\dot{p}G)-\dot{\Lambda}}
{\frac{8\pi}{c^4}(Gp)-\Lambda}+\frac{2\dot{B}}{B}\right]\xi^0\theta+c_{28}(t,r,\phi),\label{150}\\
\xi^3&=&\frac{-1}{2}\left[\frac{\frac{8\pi}{c^4}(\dot{G}p+\dot{p}G)-\dot{\Lambda}}
{\frac{8\pi}{c^4}(Gp)-\Lambda}+\frac{2\dot{B}}{B}\right]\xi^0\phi+c_{29}(t,r,\theta).\label{151}
\end{eqnarray}
The following ODEs become the necessary and sufficient conditions:
\begin{eqnarray}
\partial_t\left[\left(\frac{\frac{8\pi}{c^4}(\dot{G}p+\dot{p}G)-\dot{\Lambda}}
{\frac{8\pi}{c^4}(Gp)-\Lambda}+\frac{2\dot{A}}{A}\right)
\left(\frac{8 \pi G\rho}{c^4}+\Lambda\right)^{-1/2}\right]&=&0,\label{152}\\
\partial_t\left[\left(\frac{\frac{8\pi}{c^4}(\dot{G}p+\dot{p}G)-\dot{\Lambda}}
{\frac{8\pi}{c^4}(Gp)-\Lambda}+\frac{2\dot{B}}{B}\right)\left(\frac{8
\pi G\rho}{c^4}+\Lambda\right)^{-1/2}\right]&=&0.\label{153}
\end{eqnarray}
With the equation of state and then replacing $\frac{8\pi G
p}{c^4}=\Phi$, the above ODEs provide
\begin{eqnarray}
A&=&A_0(\Phi-\Lambda)^{-1/2}+e^{a_3\int{\sqrt{\frac{\Phi}{k}+\Lambda }dt}},\label{154}\\
B&=&B_0(\Phi-\Lambda)^{-1/2}+e^{a_4\int{\sqrt{\frac{\Phi}{k}+\Lambda}~dt}},\label{155}
\end{eqnarray}
where $A_0,~B_0,~ a_3$, and $a_4$ are constants of integration. If we
substitute these values of the metric functions in the EFEs, we get
much more complicated expressions for $G$ and $\Lambda$. However, if
we assume $\xi^0=t+c_0$, then the following relation is satisfied:
\begin{equation}\label{156}
\pounds_{HO}\left(\frac{8 \pi G(t)}{c^4}T_{ab}+\Lambda(t)
g_{ab}\right)=0,
\end{equation}
yielding the metric functions in the form
\begin{eqnarray}
A&=&A_0(t+c_0)^{\alpha_1}(\Phi-\Lambda)^{-1/2},\label{157}\\
B&=&B_0(t+c_0)^{\alpha_2}(\Phi-\Lambda)^{-1/2}.\label{158}
\end{eqnarray}
Here, $\alpha_1$ and $\alpha_2$ are constants of integration.
When we use these values of the metric functions in the EFEs, we are
again unable to discuss the behaviors of the $G$ and $\Lambda$ due to
their complicated expressions. We note that if we replace $\Lambda$
by $\Phi-1$, i.e., $\Lambda=\frac{8\pi G p}{c^4}-1$, then it also
corresponds to the homothetic case of the KSS.

It is interesting to note that if we write Eq. (\ref{104}) as
\begin{equation}\label{159}
\pounds_\xi\left(\frac{8\pi
G(t)}{c^4}T_{ab}\right)=0=\pounds_\xi({\Lambda(t)}g_{ab}),
\end{equation}
which is a special case, we may get some insight. In this way, we
reach again homothetic cases and obtain the following results:
\begin{equation}\label{160}
G \rho \approx (t+c_0)^{-2}
~~\textmd{and}~~\Lambda=\Lambda_0(t+c_0)^{-2},
\end{equation}
where $\Lambda_0$ is an arbitrary constant of integration. Further
discussion on the behavior of $\Lambda$ is the same as given in
Eq. (\ref{68}).

\begin{center}
\section*{V. SUMMARY AND DISCUSSION}
\end{center}
We have studied the perfect fluid Bianchi type III and
Kantowski-Sachs spacetimes with time-varying constants $G$ and
$\Lambda$. Due to the time-varying nature of these constants,
Bianchi identities, along with the energy-momentum conservation law,
$T^{ab}_{;b}=0$, yield a time-dependent expression of the energy
density. This expression helps us to define Hubble parameter and the
deceleration parameter.

In the KSS, we studied the behaviors of $G$ and $\Lambda$ for the
first and the second kinds. When we solve the system of ten
self-similar equations simultaneously, there arise two ODEs as
necessary and sufficient conditions. The solution of these ODEs
yields the metric functions $A$ and $B$, which make the variational
behaviors of $G$ and $\Lambda$ possible. We also discuss the dust,
radiation, and stiff fluid cases. Further, we discuss the behaviors
of $G$ and $\Lambda$ by using MCs only the case that corresponds to
the KSS.

In the KSS of the first kind, the metric functions take the form
$$A=A_0(t+c_0)^{\alpha_1}~~ \textmd{and} ~~B=B_0(t+c_0)^{\alpha_2},$$ where
$\alpha_1,\alpha_2,A_0$, and $B_0$ are arbitrary positive constants.
It is worth mentioning here that the physical situation is only
possible if we assume $\alpha_2=1$. Then, $G$ takes the form
$$G=G_0(t+c_0)^{\alpha_1+(\alpha_1+2)k},$$ where
$G_0$ is a constant given by Eq. (\ref{54}).

For a Bianchi Type III metric, $\alpha_1=\sqrt{1-\frac{c^2}{{B_0}^2}}$,
and the behavior of $G$ depends on $b$, Eq. (\ref{2.2.121a}). For
the Kantowski-Sachs metric, $\alpha_1$ turns out to be
$\sqrt{1+\frac{c^2}{{B_0}^2}}$ and the behavior of $G$ depends on
$d=\frac{c(1+k)}{\sqrt{3k^2-1-2k}}\quad \forall \quad k \in
(-\infty, -\frac{1}{3})\cup(1, \infty)$.

\begin{center}
{\bf {\small Table 1.}} {\small Behavior of $G$ for both spacetimes.}

\vspace{0.25cm}

\begin{tabular}{|l|l|l|l|}
\hline {\bf Spacetime} &{\bf $G$ is increasing} & {\bf Constant}
\\ \hline Bianchi type III &$B_0 \in \Re^+ \backslash (0,c)$ & $B_0= b$
\\ \hline Kantowski-Sachs & Any value of $B_0 \in \Re^+$ & $B_0= d$
\\ \hline
\end{tabular}
\end{center}
The dust, radiation and stiff fluid cases for Bianchi Type III
metric are given in Table $2$.
\begin{center}
{\bf {\small Table 2.}} {\small Behavior of $G$ for different cases
of fluids.}

\vspace{0.25cm}

\begin{tabular}{|l|l|l|l|}
\hline {\bf Behavior of $G$} & {\bf Dust case}& {\bf Radiation
case}& {\bf Stiff matter}
\\ \hline $\textmd{$G$ is increasing}$ &  $B_0 \in \Re^+ \backslash (0,c)$ &
  $B_0\in\Re^+ \backslash (0,\frac{2c}{\sqrt{3}})$&$B_0 \in \Re^+ \backslash (0,c)$
\\ \hline $\textmd{$G$ is Constant}$  &$B_0= c$
 &$B_0= \frac{2c}{\sqrt{3}}$ & ~~~-~~~
\\ \hline
\end{tabular}
\end{center}
In the case of the Kantowski-Sachs metric, the dust and radiation cases are
not physical while for stiff matter, $G$ is increasing $\forall~ B_0
\in \Re^+$.
In the KSS of the first kind, $\Lambda$ is given by the following
equation:
\begin{eqnarray*}
\Lambda=\Lambda_0(t+c_0)^{-2},
\end{eqnarray*}
which yields the behavior of $\Lambda$ given below in Table $3$.
In Bianchi type III  and Kantowski-Sachs metrics, the possible
values of $\Lambda_0$ are discussed in Table $4$. For various fluids,
$\Lambda_0$ reduces for Bianchi type III, as given in Table $5$.
In Kantowski-Sachs metric, the solutions corresponding to the dust and
the radiation cases are not physical while for stiff matter $\Lambda_0>0
~~\forall~~B_0 \in \Re^+$.
\begin{center}
{\bf {\small Table 3.}} {\small Behavior of $\Lambda$.}

\vspace{0.25cm}

\begin{tabular}{|l|l|}
\hline {\bf Behavior of $\Lambda$} & {\bf Value of $\Lambda_0$ and
$t$}
\\ \hline $\Lambda$ is increasing  & $\Lambda_0<0$ and $t>c_0$
 or $\Lambda_0>0$ and $t<c_0$
\\ \hline $\Lambda$ is decreasing & $\Lambda_0>0$ and $t>c_0$
 or $\Lambda_0<0$ and $t<c_0$
\\ \hline $\Lambda$ vanishes & $\Lambda_0=0$ or $t\rightarrow\infty$
\\ \hline
\end{tabular}
\end{center}

\vspace{0.25cm}

\begin{center}

\vspace{0.25cm}

{\bf {\small Table 4.}} {\small Possible values of $\Lambda_0$.}

\vspace{0.25cm}

\begin{tabular}{|l|l|l|}
\hline {\bf Spacetime} & {\bf $\Lambda_0>0$} & {\bf $\Lambda_0=0$ }
\\ \hline Bianchi type III , &$B_0 \in \Re^+ \backslash (0,c)$& $B_0= b$
\\ \hline Kantowski-Sachs& $~\forall~B_0\in \Re^+$ & $B_0= d$
\\ \hline
\end{tabular}
\end{center}

\begin{center}

\vspace{0.25cm}

{\bf {\small Table 5.}} {\small Possible values of $\Lambda_0$ for
different types of fluids.}

\vspace{0.25cm}

\begin{tabular}{|l|l|l|l|}
\hline {\bf Cases} & {\bf Dust Fluid} & {\bf Radiation Case } & {\bf
Stiff Matter}
\\ \hline $\Lambda_0>0$ & $B_0 \in \Re^+ \backslash (0,c)$
& $B_0 \in \Re^+ \backslash (0 ,\frac{2c}{\sqrt{3}})$ & $B_0 \in
\Re^+ \backslash (0,c)$
\\ \hline $\Lambda_0=0$& $B_0=c$& $B_0=
\frac{2c}{\sqrt{3}}$&~~~-~~~
\\ \hline
\end{tabular}
\end{center}
We note that for the KSS of the second kind, the results coincide with
those of the first kind as given in the above tables except that in
the expressions for the metric functions, $c_0$ is replaced with a
fraction of two constants $\frac{\beta}{\alpha}$.

Using the MCs technique, the behaviors of $G$ and $\Lambda$ are not
straightforward as in the case of the KSS due to the complicated metric
functions. However, we have managed to discuss two particular cases
depending upon $\rho$. Firstly, for $\rho=\rho_0$ (a constant), $G$
becomes constant while $\Lambda$ varies as given in Eq. (\ref{131}).
Secondly, the case for $\rho=\frac{1}{(at+b)^2}$, corresponds to the
homothetic case. In modified MCs, the behaviors of $G$ and $\Lambda$
could not be discussed generally, but we obtained a homothetic case by
assuming $G(t)\rho(t)=\frac{1}{(at+b)^2}$. Further, we obtained a
relationship $G \rho \approx (t+c_0)^{-2}$, which was the same as
obtained in the KSS technique. Similarly, in the completely modified MCs
case, we again obtained a homothetic case by assuming $\xi^0=(t+c_0)$
and $\Lambda=\frac{8\pi G p}{c^4}-1$. Further, when we re-interpreted
the completely modified MC equations, given in Eq. (\ref{104}), by
Eq. (\ref{159}), we directly obtained a homothetic case giving
$\Lambda=\Lambda_0(t+c_0)^{-2}$. Consequently, the behavior of
$\Lambda$ turns out to be the same as given in Eq. (\ref{68}) with the
KSS. It is mentioned here that the metric functions and vector
fields are the same for both spacetimes. However, we obtain
different behaviors of $G$ and $\Lambda$ due to slight changes in
the EFEs.

We found that the cosmological constant $\Lambda$ turned out to be
a time decreasing function for $\Lambda_0>0$ while the gravitational
constant $G$ was a time increasing function when $G_0>0$ for all
values of $k>0$. It is worth mentioning here that our results verify
the results obtained by Belinchon [29]. For these behaviors of $G$ and
$\Lambda$, the time-dependent vacuum energy density relation is also
satisfied, according to which both these constants are changing in a
reciprocal way [26]. Further, we found that for $\alpha_2=1$, the
deceleration parameter attained a negative value, which showed that the
expansion of the universe was accelerating. Thus, we can say that with
the expansion of the universe, $\Lambda$ is going to reduce [5].

We would like to mention here that the above mentioned time-varying
behaviors of $G$ and $\Lambda$ can only be discussed in the
homothetic case, i.e, the KSS of the first kind for both Bianchi
type III and Kantowski-Sachs spacetimes. Moreover, we found that the
vector field satisfying equation $\pounds_\xi g_{ab}=2g_{ab}$ also
satisfied equation $\pounds_\xi T_{ab}=0$. By modifying the MC
equations in an appropriate way, we were able to find the same
relationships as in the case of the KSS solution.

\vspace{0.5cm}
\begin{center}
{\bf ACKNOWLEDGMENTS}
\end{center}

\vspace{0.5cm} We would like to acknowledge Higher Education
Commission Islamabad for its financial support through the {\it
Indigenous PhD 5000 Fellowship Program Batch-III}. We appreciate the
fruitful discussions with Mr. Jamil.

\vspace{0.5cm}
\begin{center}
{\bf \large REFERENCES}
\end{center}

\begin{description}

\item{[1]} V. Sahani and A. Starobinsky, Int. J. Mod. Phys. \textbf{9}, 373(2000).

\item{[2]} A.G. Reiss, et al., Astron. J. \textbf{116}, 1009(1998).

\item{[3]} S. Perlmutter, et al., Nature \textbf{391}, 51(1998); Astrophys. J. \textbf{517}, 565(1999).

\item{[4]} M. Sharif and Z. Ahmad, Mod. Phys. Lett.
\textbf{A22}, 1493(2007).

\item{[5]} Abdussattar and R.G. Vishwakarma, Pramana J. Phys. \textbf{47}, 41(1996);
R.G. Vishwakarma, Class. Quantum Gravit. \textbf{17}, 3833(2000);
\textit{ibid} \textbf{18}, 1159(2001); \textbf{19}, 4747(2002); Gen.
Relat. Gravit. \textbf{33}, 1973(2001); Mon. Not. Roy. Astron. Soc.
\textbf{331}, 776(2002) and references therein.

\item{[6]} J. Kokosar, \textit{The Variable Gravitational Constant
G, General Relativity Theory, Elementary Particles, Quantum
Mechanics, Time's Arrow and Consciousness}, PHILICA.COM, Article number 17(2006).

\item{[7]} P.A.M. Dirac, Nature \textbf{139}, 323(1937).

\item{[8]} P. Lore-Aguilar, E. Garci-Berro, J. Isern and A.Kubyshin,
Class. Quantum Gravit. \textbf{20}, 3885(2003).

\item{[9]} I.I. Shapiro, W.B. Smith, M.D. Ash, R.P. Ingalls and G.H.
Pettengill, Phys. Rev. Lett. \textbf{26}, 27(1971).

\item{[10]} R.H. Dicke, Science \textbf{138}, 635(1962).

\item{[11]} C. Brans and R.H. Dicke, Phys. Rev. \textbf{124}, 925(1961).

\item{[12]} C.J. Copi,  A.N. Davis and L.M. Krauss, Phys. Rev. Lett. \textbf{92}, 171301(2004).

\item{[13]} J.D., Bekenstien, Found. Phys. Rev. \textbf{16}, 409(1986).

\item{[14]} O. Bertolami, Nuovo Cimento \textbf{93}, 36(1986).

\item{[15]} A-M.M. Abdel-Rehman, Nuovo Cimento \textbf{B102}, 225(1988);
Phys. Rev. \textbf{D45}, 3497(1992).

\item{[16]} M.S. Berman, Phys. Rev. \textbf{D43}, 1075(1991).

\item{[17]} J.C. Carvalho, J.A.S., Lima and I. Waga, Phys. Rev. \textbf{D46}, 2404(1992).

\item{[18]} B. Saha, Astrophys. Space Sc. \textbf{302}, 83(2006).

\item{[19]} A. Beesham, Gen. Relativ. Gravit. \textbf{26}, 159(1994).

\item{[20]} J.P. Singh, A. Pradhan and A.K. Singh, Astrophys. Space Sc. \textbf{314}, 83(2008).

\item{[21]} R.G. Vishwakarma, Gen. Relativ. Gravit. \textbf{37}, 1305(2005).

\item{[22]} A.I. Arbab, Gen. Relativ. Gravit. \textbf{30}, 1401(1998).

\item{[23]} A. Pradhan and P. Pandey,  Astrophys. Space Sc. \textbf{301}, 127(2006).

\item{[24]} B. Saha, Mod. Phys. Lett. \textbf{A6}, 1287(2001).

\item{[25]} D. Kalligas, P.S. Wesson and C.W.F. Everitt,  Gen. Relativ.
Gravit. \textbf{27}, 645(1995).

\item{[26]} F. Darabi, \textit{Time Variation of $G$ and
$\Lambda$, Acceleration of the Universe, Coincidence problem and
Mach's Cosmological Coincidence }, gr-qc/0802.0028.

\item{[27]} J.A. Belinchon and P. Dávila,
Class. Quantum Grav. \textbf{17}, 3183(2000); \textit{An Eexcuse for
Revising a Theory of Time Varying Constants}, gr-qc/0404028.

\item{[28]} J.A. Belinchon,
Gravitation and Cosmology \textbf{15}, 306(2009) .

\item{[29]} J.A. Belinchon, Astrophys. Space Sc. \textbf{315}, 111(2008).

\item{[30]} M.A.H. MacCallum, \textit{In General Relativity:
An Einstien Centenary Survey,} eds. Hawking, S.
 and Israel, W. (Cambridge Univ. Press, 1979)533.

\item{[31]} M.E. Cahill and A.H. Taub, Commun. Math. Phys. \textbf{21}, 1(1971).

\item{[32]} B. Carter and R.N. Henriksen, Annales De Physique \textbf{14}, 47(1989);
J. Math. Phys. \textbf{32}, 2580(1991).

\item{[33]} M. Sharif and S. Aziz, Class. Quantum Gravit. \textbf{24}, 605(2007).

\end{description}
\end{document}